\documentclass[epj]{svjour}
\usepackage{color,graphicx}
\usepackage{graphics}
\usepackage{amsmath}
\begin{document}
\title{Low-lying states in even Gd isotopes studied with five-dimensional collective
Hamiltonian based on covariant density functional theory}
%\subtitle{Do you have a subtitle?\\ If so, write it here}
\author{Z. Shi\inst{1} \and Q. B. Chen\inst{2,3}
%\thanks{\emph{Present address:} Physik-Department, Technische Universit\"{a}t M\"{u}chen, D-85747 Garching, Germany}
\and S. Q. Zhang\inst{2,}\thanks{sqzhang@pku.edu.cn}% etc
% \thanks is optional - remove next line if not needed
%
}                     % Do not remove
%
%\offprints{}          % Insert a name or remove this line
%
\institute{School of Physics and Nuclear Energy Engineering,
Beihang University, Beijing 100191, China
\and State Key Laboratory of
Nuclear Physics and Technology, School of Physics, Peking
University, Beijing 100871, China
\and Physik-Department, Technische Universit\"{a}t M\"{u}nchen, D-85747 Garching, Germany}
\date{Received: date / Revised version: date}
% The correct dates will be entered by Springer
%
\abstract{
Five-dimensional collective Hamiltonian based on the covariant density functional
theory has been applied to study the the low-lying states of even-even $^{148-162}$Gd
isotopes. The shape evolution from $^{148}$Gd to $^{162}$Gd is presented. The experimental
energy spectra and intraband $B(E2)$ transition probabilities for the $^{148-162}$Gd
isotopes are reproduced by the present calculations. The relative $B(E2)$ ratios in
present calculations are also compared with the available interacting boson model 
results and experimental data. It is found that the occupations of neutron $1i_{13/2}$
orbital result in the well-deformed prolate shape, and are essential for Gd isotopes.
 % end of PACS codes
} %end of abstract
\maketitle
\section{Introduction}
\label{intro}

The rare-earth region, which possesses many transitional and well-deformed nuclei,
is an ideal venue to study the origination of deformation and collective motion. One
interesting and important example is the Gadolinium isotopes, the structures of which
have attracted a large number of experimental and theoretical studies.

A well-known phenomenon in Gd isotopes is the presence of quantum phase transition.
As pointed in Refs.~\cite{casten2001empirical,tonev2004transition}, $^{154}$Gd locates
at the X(5) critical point of a vibrator to axial rotor phase transition. The quantum
phase transition in the Gd isotopes has been investigated by the potential energy surfaces
(PESs) within the framework of microscopic mean-field theory. For the relativistic
mean-field (RMF) calculations, the shape transition between spherical and axially deformed
nuclei in rare-earth region is investigated by the microscopic quadrupole constrained RMF
theory with the NL3 interaction~\cite{sheng2005systematic}, and the possible critical point
nuclei are suggested to be $^{150, 152}$Gd. For the results obtained from the relativistic
Hartree-Bogoliubov (RHB) calculations with NL3~\cite{fossion2006e5}, a relatively
flat PES occurs for the $^{150}$Gd. Whereas, the PES of $^{154}$Gd is not flat and exhibits
a deeper minimum in the prolate region and a shallower minimum in the oblate region. For
the non-relativistic case, a self-consistent Skyrme-Hartree-Fock plus BCS calculations is
performed~\cite{guzman2007e5}. In this case, the transitional behavior for $^{152, 154}$Gd
is found, and the flat behavior, which is one of the expected characteristics of X(5) critical
point symmetries, is not found in the PESs of the $N = 90$ isotones. This is in agreement
with relativistic mean-filed calculation~\cite{fossion2006e5}. Another non-relativistic
mean-field calculations was performed with self-consistent Hartree-Fock-Bogoliubov approximation
based on the finite-range and density-dependent Gogny interaction~\cite{robledo2008evolution},
including $^{152,154}$Gd. In the case of X(5) candidates, a shape transition from spherical
to well-deformed nuclei was found, but the realization of the X(5) symmetry in terms of
flat (quadratic) patterns in the $\beta_2$ ($\gamma$) potentials is uncertain~\cite{robledo2008evolution}.

Another interesting investigation in the Gd isotopes is the so-called partial dynamical symmetry
(PDS)~\cite{leviatan1996partial,van1999dynamical,leviatan2002generalized,leviatan2011partial}.
In Ref.~\cite{leviatan2013partial}, the SU(3) PDS was proposed to be used as a selection
criterion for higher order terms in the general interacting boson model (IBM) Hamiltonian.
Taking $^{156}$Gd as an example, the SU(3) PDS improves the description of odd-even staggering
of the $\gamma$ band, which is absent in the SU(3) dynamical symmetry. Systematic examinations
of SU(3) PDS in finite nuclei have been recently carried out in Ref.~\cite{casten2014evidence}.
The relative interband $B(E2)$ ratios of 47 deformed nuclei in the rare-earth region were
first compared with the available experimental data~\cite{casten2014evidence}, including
$^{158}$Gd. Besides, the link between the concepts of PDS and quasi-dynamical symmetry
(QDS)~\cite{bahri20003su,rowe2004phase} was revealed by Kremer $et~al.$ in the ground-state
band of $^{160}$Gd~\cite{kremer2014linking,kremer2015linking}. Such link has been further
demonstrated by Van Isacker~\cite{van2014partial}.

The microscopic density functional theory (DFT), which starts from an effective nucleon-nucleon
interaction and self-consistently determines the nuclear mean-field by all the independent
particles inside, has achieved a lot of successes in describing both the nuclear ground
state and excited state properties
\cite{ring1996relativistic,bender2003self,vretenar2005relativistic,meng2006relativistic,meng2015relativistic}.
Based on the DFT or its covariant version CDFT, many properties of collective structures of
nuclei are well described. In recent years, five-dimensional collective Hamiltonian based
on the covariant density functional theory (5DCH-CDFT)~\cite{nikvsic2009beyond,nikvsic2011relativistic}
has been developed and extensively applied to describe the collective properties such as the phase transitions
~\cite{li2009microscopic,li2009microscopic02,li2010microscopic,song2011microscopic,li2013simultaneous}, shape evolutions
~\cite{li2011energy,xiang2012covariant,sun2014spectroscopy,nikvsic2014microscopic,wang2015covariant,wang2016tidal}
as well as the low-lying spectra along with the isotopic and isotonic chains in different mass regions
~\cite{nikvsic2009beyond,li2010relativistic,li2012enhanced,fu2013beyond,shi2015description}.
The even $^{152-160}$Gd isotopes have been investigated by the 5DCH-CDFT with PC-F1
parameter~\cite{nikvsic2009beyond}.

In the present paper, we will reinvestigate the $^{148-162}$Gd isotopes in the framework
of 5DCH-CDFT with the density functional PC-PK1~\cite{shi2015description,zhao2010new}. The purpose
of this work is twofold: (i) By starting from $N=84$, the low-lying states for the Gd
isotopes will be discussed from the 5DCH-CDFT perspective, and a complete structure evolution
from near spherical to well deformed nuclei will be clearly illustrated. (ii) The relative
$B(E2)$ ratio, which is an observable closely related to the PDS, will be investigated.

The paper is organized as follows. In Sec.~\ref{sec1}, a brief introduction to the framework
of 5DCH is given. In Sec.~\ref{sec2}, the potential energy surfaces, excitation spectra,
intraband $B(E2)$ transition probabilities and relative $B(E2)$ ratios etc., for even-even
$^{148-162}$Gd isotopes obtained from the 5DCH-CDFT are presented. Finally, a summary is given
in Sec.~\ref{sec3}.

\section{Theoretical Framework}\label{sec1}

A detailed formalism of the 5DCH has been presented in numerous articles in literatures, see, e.g.
Refs.~\cite{prochniak1999collective,nikvsic2009beyond}. In this subsection, for completeness,
a brief introduction is presented.

The five-dimensional collective Hamiltonian, which could simultaneously treat the quadrupole
vibrational and rotational excitations, is expressed in terms of the two deformation parameters
$\beta$ and $\gamma$, and three Euler angles $(\phi,\theta,\psi)\equiv \Omega$ that define
the orientation of the intrinsic principal axes in the laboratory frame,
%-------------------------------------------------------------------------------
\begin{equation}
    \hat H_{\rm coll}(\beta,\gamma)=\hat T_{\rm vib}(\beta,\gamma)
    +\hat T_{\rm rot}(\beta,\gamma,\Omega)+V_{\rm coll}(\beta,\gamma).\label{eq1}
\end{equation}
%-------------------------------------------------------------------------------
The three terms in $\hat H_{\rm coll}(\beta,\gamma)$ are respectively the vibrational kinetic energy
%-------------------------------------------------------------------------------
\begin{equation}
\begin{split}
\hat{T}_{\rm vib}=&-\frac{\hbar^2}{2\sqrt{wr}}\Bigg\{\frac{1}{\beta^4}\Big
[\frac{\partial}{\partial \beta}\sqrt{\frac{r}{w}}\beta^4B_{\gamma\gamma}\\
&-\frac{\partial}{\partial\beta}\sqrt{\frac{r}{w}}\beta^3 B_{\beta\gamma}
\frac{\partial}{\partial\gamma}\Big]
+\frac{1}{\beta \sin 3\gamma}\Big[-\frac{\partial}{\partial\gamma}\\
&\times \sqrt{\frac{r}{w}}\sin3\gamma B_{\beta\gamma}\frac{\partial}{\partial\beta}+
\frac{1}{\beta}\frac{\partial}{\partial\gamma}\sqrt{\frac{r}{w}}
\sin 3\gamma B_{\beta\beta}\frac{\partial}{\partial\gamma}\Big]\Bigg\},
\label{eq2}
\end{split}
\end{equation}
%-------------------------------------------------------------------------------
the rotational kinetic energy
%-------------------------------------------------------------------------------
\begin{equation}
   \hat T_{\rm rot}=\frac{1}{2}\sum^3_{k=1}\frac{\hat J_k^2}{\mathcal I_k},
   \label{eq3}
\end{equation}
%---------------------------------------------------------------------------------
and the collective potential $V_{\rm coll}$. $\hat J_k$ $(k=1,~2,~3,)$ denote the components
of the angular momentum in the body-fixed frame of a nucleus, and the mass parameters
$B_{\beta\beta}$, $B_{\beta\gamma}$, $B_{\gamma\gamma}$, as well as the moments of inertia
$\mathcal I_k$ depend on the quadrupole deformation variables $\beta$ and $\gamma$. Two
additional quantities that appear in the $\hat{T}_{\rm vib}$ term~(\ref{eq2}), $r = B_1B_2B_3$
and $w = B_{\beta\beta} B_{\gamma\gamma} - B_{\beta\gamma}^2$, determine the volume element
in the collective space.

The eigenvalue problem of the Hamiltonian~(\ref{eq1}) is solved using an expansion of
eigenfunctions in terms of a complete set of basis functions that depend on the five
collective coordinates $\beta, \gamma$ and $\Omega~(\phi,\theta,\psi)$~\cite{nikvsic2009beyond}.
Using the obtained collective wave functions
\begin{align}
    \Psi^{IM}_\alpha(\beta, \gamma, \Omega)
    =\sum_{K\in\Delta I}\psi^I_{\alpha K}(\beta, \gamma)\Phi^I_{MK}(\Omega),
\end{align}
various observables such as electric quadrupole transition probabilities can be calculated,
\begin{align}
    B(E2;\alpha I\rightarrow\alpha'I')
    =\frac{1}{2I+1}|\langle\alpha'I'||\hat M(E2)||\alpha I\rangle|^2,
\end{align}
where $\hat M (E2)$ is the electric quadrupole operator~\cite{nikvsic2009beyond}.

In the framework of 5DCH-CDFT, the microscopic collective parameters of 5DCH are all
determined from the CDFT, which include the mass parameters $B_{\beta\beta}$,
$B_{\beta\gamma}$, and $B_{\gamma\gamma}$, the moments of inertia $\mathcal I_k$, and
the collective potential $V_{\rm coll}$. The moments of inertia are calculated with the
Inglis-Belyaev formula and the mass parameters from the cranking approximation.
$V_{\rm coll}$ is obtained by subtracting the zero-point energy corrections from the
total energy that corresponds to the solution of constrained triaxial CDFT. Detailed
formalism can be found in Ref.~\cite{nikvsic2009beyond}.

%-------------------------------------------------------------------------------
\begin{figure*}[t!]
  \centering
  \includegraphics[scale=0.7,angle=0]{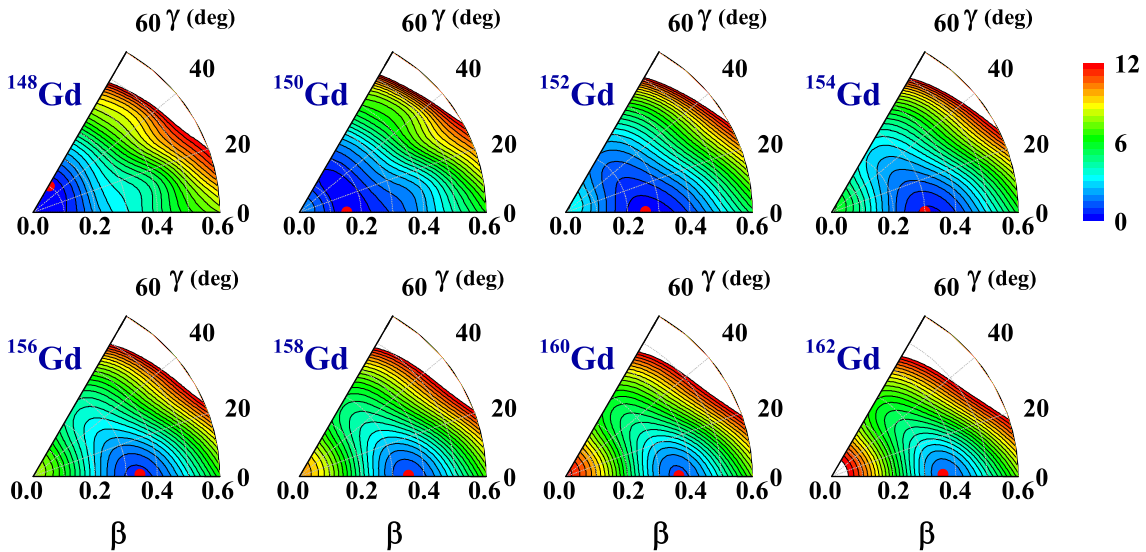}
  \caption{(Color online) The potential energy surfaces of even-even $^{148-162}$Gd isotopes
  in the $\beta$-$\gamma$ planes calculated by constrained triaxial covariant density functional
  theory with PC-PK1 \cite{zhao2010new}. All energies are normalized with respect to the binding
  energy of the absolute minimum (labeled by a red dot). The energy difference between the
  neighbouring contour lines is 0.5 MeV.}
\label{PES-Gd}
\end{figure*}
%-------------------------------------------------------------------------------

\section{Results and discussion}\label{sec2}

The main purpose of this study is a microscopic analysis for the collective potential
energy surfaces, low-energy collective spectra, relative $B(E2)$ transition ratios
etc., in the even $^{148-162}$Gd isotopes using the CDFT-based 5DCH model. To
determine the collective parameters for the 5DCH, we perform a constrained RMF
plus BCS (RMF+BCS) calculation, with the effective interaction in the particle-hole
channel defined by the point-coupling density functional PC-PK1~\cite{zhao2010new},
and a density-independent $\delta$-force in the particle-particle channel. The strength
parameter of the $\delta$-force is 349.5 MeV fm$^3$ (330.0 MeV fm$^3$) for neutron
(proton), which is determined by fitting the empirical neutron (proton) pairing
gap~\cite{zhao2010new}. The Dirac equation is solved by expanding the Dirac spinor
in terms of the three dimensional harmonic oscillator basis with 14 major shells.

\subsection{Potential energy surfaces}

Firstly, to get a comprehensive understanding of the shape evolution in Gd isotopes,
the PESs of even-even $^{148-162}$Gd isotopes, calculated by constrained triaxial
CDFT with the PC-PK1 \cite{zhao2010new} density functional are shown in Fig.~\ref{PES-Gd}
in the $\beta$-$\gamma$ planes.

For $^{148}$Gd, the minimum of the PES locates at $\beta=0.1$ and $\gamma=60^\circ$,
indicating that the ground state of $^{148}$Gd is of slight oblate deformation. Meanwhile,
the potential energy in the region of $|\beta|\leq0.1$ exhibits a rather weak dependence
on both $\beta$ and $\gamma$. With two more neutrons, the ground state of $^{150}$Gd
is prolate with $\beta=0.15$, and its PES around the ground state is extremely flat
in the $\gamma$ direction, where a tunnel connecting the prolate and oblate shapes can
be seen. With further more neutrons, the quadrupole deformation $\beta$ of Gd isotopes
increases from $\beta=0.15~{\rm (^{150}Gd)~to}~0.25$ ($^{152}$Gd), to 0.3 ($^{154}$Gd),
and to $0.35$ ($^{156}$Gd), with the triaxial deformation parameter keeping at
$\gamma = 0^\circ$. Continuing to add neutrons, the deformations of the ground states
of $^{158,160,162}$Gd become the same as that of $^{156}$Gd, i.e., $\beta=0.35$ and
$\gamma=0^\circ$. It is noted that the potential rigidness along both $\beta$ and
$\gamma$ directions become more and more rigid with the increase of the neutron number.

Thus, a clear shape evolution from weakly deformed ($^{148,150}$Gd) to soft prolate
($^{152,154}$Gd) and to the well deformed prolate ($^{156-162}$Gd) for the even-even
$^{148-162}$Gd isotopes is presented.

The PESs shown in Fig.~\ref{PES-Gd}, including the locations of the minima, are consistent
with the calculations with PC-F1 parameter~\cite{nikvsic2009beyond}, except that the PESs
with PC-PK1 parameter are slightly rigid in the $\gamma$ direction. In addition, 
the PES of $^{154}$Gd in the present calculation does not exhibit the flat behavior.
This is consistent with the previous mean-field
calculations~\cite{fossion2006e5,guzman2007e5,robledo2008evolution}.

%-------------------------------------------------------------------------------
\subsection{Collective energy spectra}
%-------------------------------------------------------------------------------

%-------------------------------------------------------------------------------
\begin{figure*}[ht!]
  \centering
  \includegraphics[scale=0.42,angle=0]{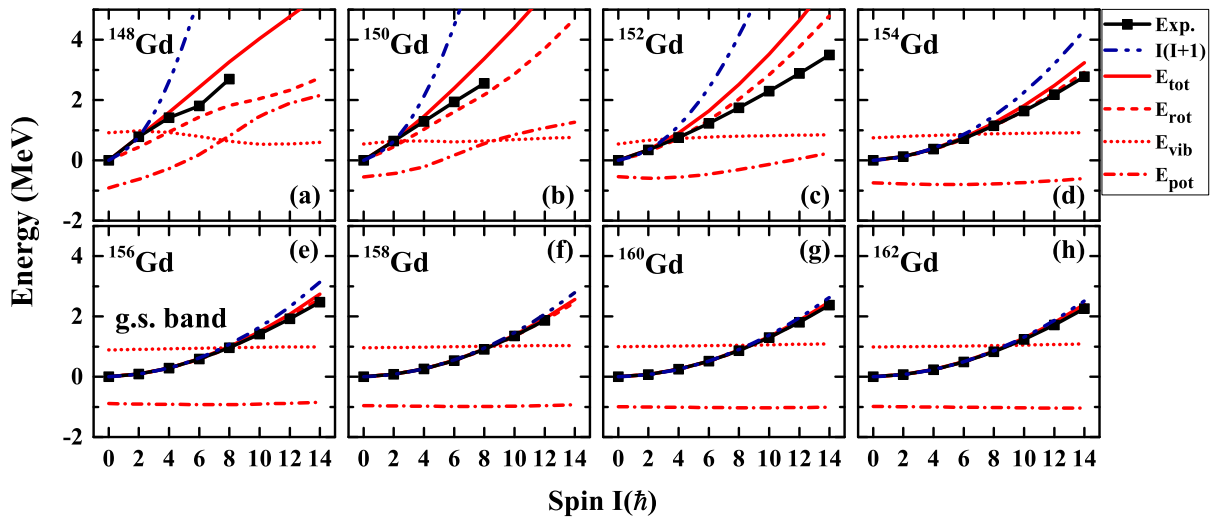}
  \caption{(Color online) The energy spectra for ground state bands in even-even $^{148-162}$Gd
  isotopes calculated by 5DCH model based on CDFT in comparison with those available experimental
  data. The theoretical spectra $E_{\rm tot}$ are normalized to the experimental energy of
  $2^+_1$. $E_{\rm rot}$, $E_{\rm vib}$ and $E_{\rm pot}$ represent the energy contributions
  from the rotational, vibrational and collective potential energy terms in the collective
  Hamiltonian. All the experimental data are taken from NNDC~\cite{http://www.nndc.bnl.gov/}.}
\label{spectra-gsband}
\end{figure*}
%-------------------------------------------------------------------------------

\begin{figure*}[t!]
  \centering
  \includegraphics[scale=0.43,angle=0]{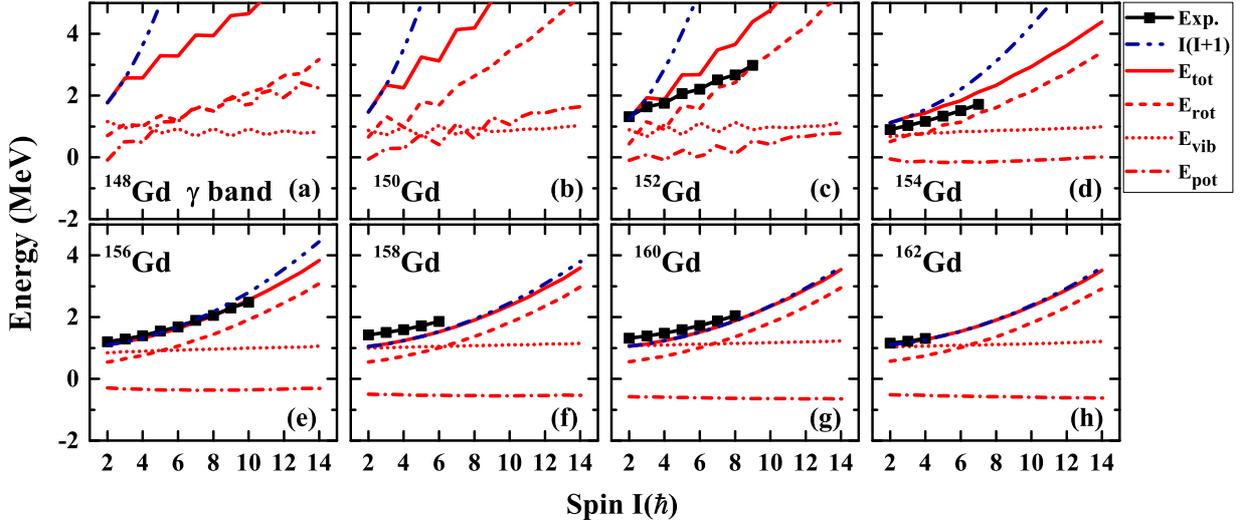}
  \caption{(Color online) The energy spectra for $\gamma$ bands in even-even $^{148-162}$Gd
  isotopes   calculated by 5DCH model based on CDFT in comparison with those available
  experimental data. $E_{\rm rot}$, $E_{\rm vib}$ and $E_{\rm pot}$ represent the energy
  contributions from the rotational, vibrational and collective potential energy terms in
  the collective Hamiltonian. All the   experimental data are taken from
  NNDC~\cite{http://www.nndc.bnl.gov/}.}
\label{spectra-gammaband}
\end{figure*}

\begin{figure*}[t!]
  \centering
  \includegraphics[scale=0.42,angle=0]{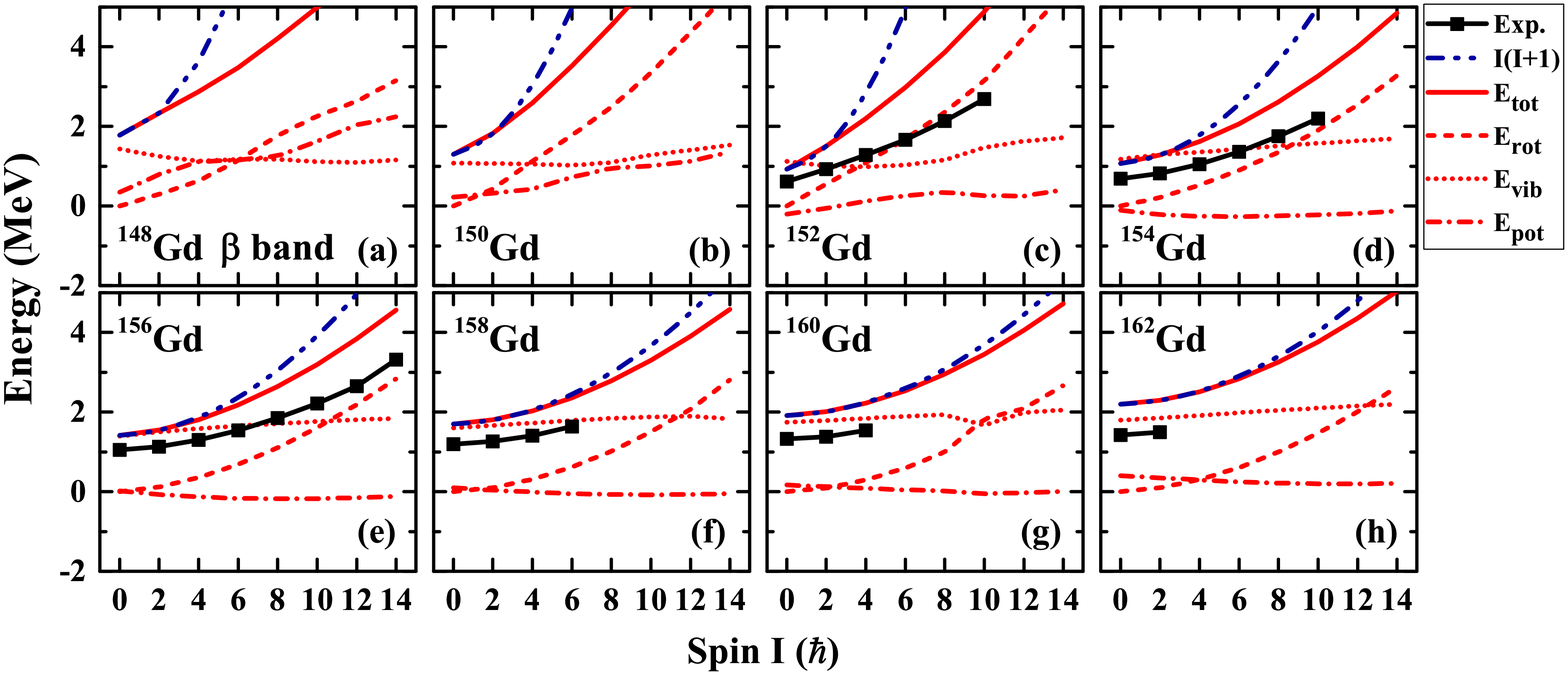}
  \caption{(Color online) The energy spectra for $\beta$ bands in even-even $^{148-162}$Gd
  isotopes   calculated by 5DCH model based on CDFT in comparison with those available
  experimental data. $E_{\rm rot}$, $E_{\rm vib}$ and $E_{\rm pot}$ represent the energy
  contributions from the rotational, vibrational and collective potential energy terms
  in the collective Hamiltonian. All the experimental data are taken from
  NNDC~\cite{http://www.nndc.bnl.gov/}.}
\label{spectra-betaband}
\end{figure*}

With the collective parameters determined from CDFT, the excitation energies and the collective
wave functions for each value of the total angular momentum $I$ can be obtained by diagonalizing
the 5DCH Hamiltonian. As discussed in Ref.~\cite{nikvsic2009beyond}, in the 5DCH results, the
$\gamma$ and $\beta$ bands in the deformed and transitional nuclei are usually assigned according
to the distribution of the angular momentum projection $K$. That is the excited states with
predominant $K=2$ components in the wave function are assigned to be $\gamma$ band, whereas the
states above the yrast sequence characterized by dominant $K=0$ components are assigned to be
$\beta$ band. In Figs.~\ref{spectra-gsband},~\ref{spectra-gammaband}, and \ref{spectra-betaband},
the low-lying excitation energy spectra for the ground state bands, $\gamma$ bands and $\beta$
bands of Gd isotopes calculated by 5DCH are displayed, respectively. The available experimental
data~\cite{http://www.nndc.bnl.gov/} are also shown for comparison.

It is noted that currently, the inertia parameters are calculated by the Inglis-Belyaev formula,
which does not include the Thouless-Valatin dynamical rearrangement contributions and, thus would
systematically underestimate the empirical values remarkably. As illustrated in Ref.~\cite{li2012efficient},
the Thouless-Valatin corrections are almost independent of deformation and for a given nucleus
the effective moment of inertia used in the collective Hamiltonian can simply be obtained by
renormalizing the Inglis-Belyaev values by a constant factor. In the practical calculation, the
theoretical result of the $2^+_1$ state is normalized to the corresponding experimental $2^+_1$
energy. For the even-even Gd isotopes from $^{148}$Gd to $^{162}$Gd, the factors are 1.889, 0.863,
0.818, 1.379, 1.500, 1.500,1.455 and 1.411, respectively.

In Fig.~\ref{spectra-gsband}, the energies of the ground-state (g.s.) bands in Gd isotopes
obtained from 5DCH calculations are compared with the available experimental data. The energy
contribution from each term of the collective Hamiltonian, i.e., the rotational, vibrational
and collective potential energy terms are also present, which are calculated through the
following form
\begin{align}
    E_{\rm rot} &= \langle\Psi_\alpha^I|\hat T_{\rm rot}|\Psi_\alpha^I\rangle, \\
    E_{\rm vib} &= \langle\Psi_\alpha^I|\hat T_{\rm vib}|\Psi_\alpha^I\rangle, \\
    E_{\rm pot} &= \langle\Psi_\alpha^I| V_{\rm coll}|\Psi_\alpha^I\rangle.
\end{align}

In Fig.~\ref{spectra-gsband}, to illustrate the deviation of $E_{\rm tot}$ from the ideal SU(3)
character, a line calculated by $c*I(I+1)$ form is also included, in which the coefficient $c$ is
determined by the $E(2_1^+)$. As shown in Fig.~\ref{spectra-gsband}, the agreements between the
theoretical energy spectra and the experimental data for Gd isotopes are overall good. For the
light $^{148,150}$Gd isotopes, the theoretical results predict larger energy spacing. It is
noted that in the experimental ground state band of $^{148}$Gd, the energy spacing decreases with
spin when $I\leq6\hbar$. As indicated in Refs.~\cite{podolyak2000multiple,podolyak2003lifetime},
the yrast $0^+$, $2^+$, $4^+$, and $6^+$ states in $^{148}$Gd are formed by the coupling of the
two $1f_{7/2}$ neutron, which does not belong to the collective excitation and is beyond the
scope of our description. Besides, $^{148}$Gd is almost a doubly magic nucleus, in which the
neutron number is close to the magic number $N=82$ and the proton number $Z=64$ is also sometimes
regarded as a kind of magic number, and this is the main reason why the shell-model approach gives
better results. The energy spectra for both the experimental and theoretical ground
state bands deviate significantly from the ideal rigid rotor. The energy contributions
$E_{\rm rot}$, $E_{\rm vib}$ and $E_{\rm pot}$ indicate that the increase of energy with spin
comes from both the rotational and collective potential terms. This could be understood that
due to the soft PES, the deformation of $^{148,150}$Gd will increase significantly with the
increase of angular momentum, which is similar to the picture of `tidal wave'~\cite{wang2016tidal}.
For $^{152, 154}$Gd, the theoretical results overestimate the experimental data a little.
However, the trends of the level spacing increasing slightly with spin in the experimental
ground state bands are still reproduced by the calculations. For $^{156-162}$Gd, the experimental
data are well reproduced by the 5DCH calculations, both of the experimental data and theoretical
values obey the rule of $I(I+1)$ approximately, indicating that the ground state bands have
the SU(3) character. Due to the stable shapes of $^{156-162}$Gd, the increase of the ground
state bands energies all come from the rotational term, while $E_{\rm vib}$ and $E_{\rm pot}$
keep constant with the increase of spin, as can be seen in Fig.~\ref{spectra-gsband}.

%-------------------------------------------------------------------------------
%-------------------------------------------------------------------------------

%-------------------------------------------------------------------------------
\begin{figure}[htbp]
  \centering
  \includegraphics[scale=0.7,angle=0]{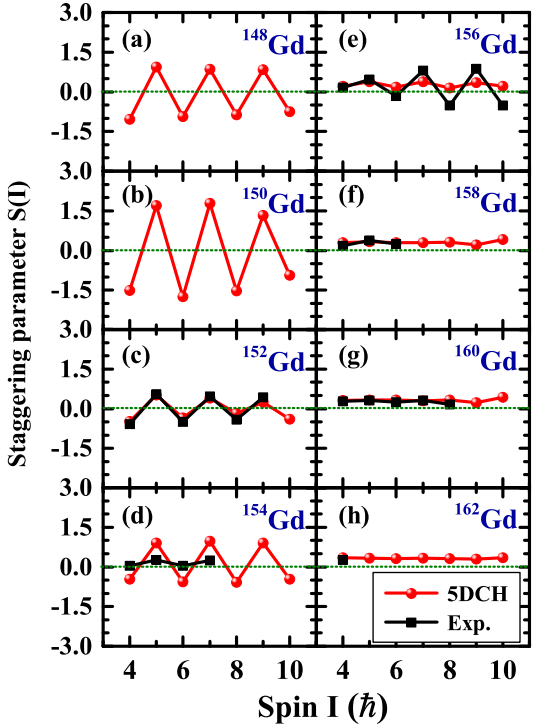}
  \caption{(Color online) The staggering parameter $S(I)$ of even-even $^{148-162}$Gd isotopes
  calculated   by 5DCH-CDFT in comparison with the experimental data~\cite{http://www.nndc.bnl.gov/}. }
\label{staggering}
\end{figure}
%-------------------------------------------------------------------------------

For the $\gamma$ bands shown in Fig.~\ref{spectra-gammaband}, the agreements between the
theoretical results and available experimental data are good. For $^{152}$Gd, the energy
difference of the bandhead between the experimental and theoretical values is about 350 keV,
and for all the other nuclei the differences are smaller than 150 keV. The increasing behaviors
for the $\gamma$ bands in $^{156-162}$Gd are consistent with the $I(I+1)$ lines, indicating
that the SU(3) symmetries are still conserved.

The $\gamma$ band staggering parameter $$S(I) = \frac{[E(I)-E(I-1)]-[E(I-1)-E(I-2)]}{ E(2^+_1 )}$$
is an indicator of the triaxial softness/rigidness \cite{zamfir1991signatures,McCutchan2007staggering}.
For a vibrator, the $S(I)$ oscillates between negative values for even-spin states and
positive values for odd-spin states, with the magnitude keeps at $|S(I)=1|$. For a nucleus
with a deformed $\gamma$-soft potential, the $S(I)$ oscillates between negative values for
even-spin states and positive values for odd-spin states, with the magnitude slowly increasing
with spin. In the limit of an axially symmetric rotor, the $S(I)$ values are positive,
small (0.33), and constant as a function of spin.

In Fig.~\ref{staggering}, we plot the theoretical staggering parameters $S(I)$ for the
$\gamma$ bands in Gd isotopes, in comparison with the available experimental data. Except
$^{154}$Gd and $^{156}$Gd, the experimental staggering parameters are well reproduced by
the 5DCH calculations. For $^{148}$Gd, the $S(I)$ parameters are negative for even-spin states and
positive for odd-spin states and the magnitude of $S(I)$ keeping about one, indicating that
$^{148}$Gd possesses the spherical vibrator behavior. For $^{150, 152,154}$Gd, the theoretical
$S(I)$ parameters are negative for even-spin states and positive for odd-spin states, close
to that of a $\gamma$ soft potential, and is in consistent with the $\gamma$ soft PESs shown
in Fig.~\ref{PES-Gd}. For $^{156-162}$Gd, the theoretical $S(I)$ parameters are close to
a constant value of 0.33, indicating the axial rotor character. The difference between the
theoretical values and experimental data in $^{154}$Gd may be due to the too soft potential
predicted by CDFT, while the difference in $^{156}$Gd may be due to the too rigid potential
predicted by CDFT.

%-------------------------------------------------------------------------------
%-------------------------------------------------------------------------------

The rotational band built on the $0_2^+$ state in deformed rare-earth nuclei has been usually
labelled as $\beta$-band for many years. In the present work, we also follow such convention.
But it is worth mentioning that the nature of the lowest-lying excited $0_2^+$ state is still
debated heavily. The investigation by Garrett $et~al.$ shows that most of the $0_2^+$ states are
not $\beta$ vibrations \cite{garrett1996nature,garrett2001characterization}. For the $\beta$
bands, as can be seen in Fig.~\ref{spectra-betaband}, the calculations do not reproduce the
data as well as the ground state bands and the $\gamma$ bands. The calculated bandhead energies
of $\beta$ band $0^+_2$ are always higher than the data. This could be understood by the fact that the
bandhead energy of the $\beta$ band depends on the mass parameters rather sensitively. In the
present calculations, the mass parameters are underestimated due to the absence of Thouless-Valatin
dynamical rearrangement contributions. One may adjust the mass parameters to reproduce the experimental
values. However, the mass parameters are so complicated, that there are no simple estimates of
the Thouless-Valatin corrections can be obtained~\cite{prochniak2004self}. Furthermore, the current
model does not contain the coupling of nuclear shape oscillations with pairing
vibrations~\cite{pilat1993coupling}. The excitation energy of the $0^+$ is also very sensitive
to such coupling~\cite{li2010microscopic}. Here, we would like to mention that, the energies
of the ground state band and $\gamma$ band do not evidently depend on the mass parameters.
Therefore, despite the lack of pairing vibration coupling in the mass parameters, the g.s. and
$\gamma$ bands can still be well reproduced. Although the $\beta$ band energies are overestimated
by the calculations, the experimental feature for the $\beta$ bands can be still reproduced.
That is the experimental and theoretical $\beta$ bands both behave as rotational bands.
%-------------------------------------------------------------------------------
\begin{figure}[htbp]
  \centering
  \includegraphics[scale=0.7,angle=0]{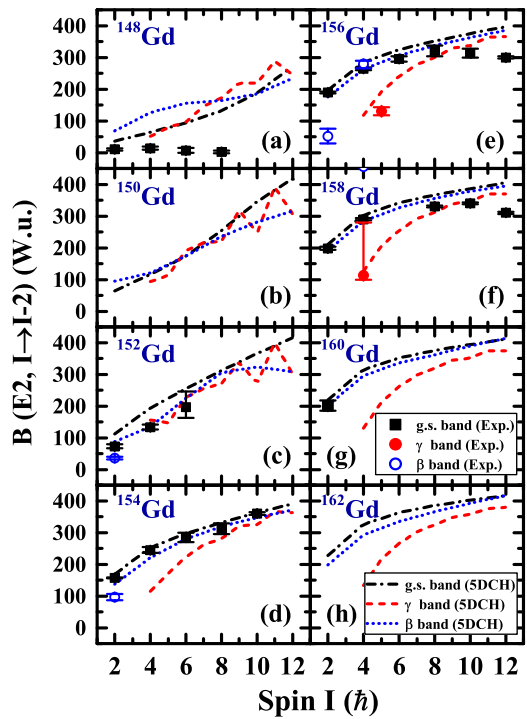}
  \caption{(Color online) The intraband $B(E2; I\to I-2)$ values (in Weisskopf unit) for the
  ground-state, $\gamma$ and $\beta$ bands in $^{148-162}$Gd. Theoretical values calculated
  with the PC-PK1 relativistic density functional are compared with the
  data~\cite{http://www.nndc.bnl.gov/}.}
\label{be2}
\end{figure}
%-------------------------------------------------------------------------------
%

For $^{156-162}$Gd, the SU(3) characters for the $\beta$ bands are a little different from
those in the g.s. and $\gamma$ bands. Generally, the $E_{\rm tot}$ of $\beta$ bands in the
$^{156-162}$Gd are consistent with the $I(I+1)$ lines in the region of low angular momentum.
With the increase of the angular momentum, the deviations of $E_{\rm tot}$ from $I(I+1)$ become
more and more apparent. This indicates the break of SU(3) symmetries in the $\beta$ bands.
The deviations with increasing angular momentum become smaller when the neutron number increases.
The increase of the total energies comes from the nuclear rotation, while the energy contributions
coming from the vibration and collective Hamiltonian remain almost unchanged, due to the well-deformed
shapes of $^{156-162}$Gd as shown in Fig.~\ref{PES-Gd}.

For the energy spectra obtained from PC-F1, the increasing trend with spin for the g.s.,
$\gamma$ and $\beta$ bands are similar to those from PC-PK1, but the bandhead energies for
the $\beta$ bands from PC-F1 are about 250 keV lower for $^{156-160}$Gd, due to the softer
PESs predicted by PC-F1~\cite{nikvsic2009beyond}.

%-------------------------------------------------------------------------------
\subsection{$B(E2)$ transition strength}
%-------------------------------------------------------------------------------

%-------------------------------------------------------------------------------
\begin{figure*}[t!]
  \centering
  \includegraphics[scale=0.4,angle=0]{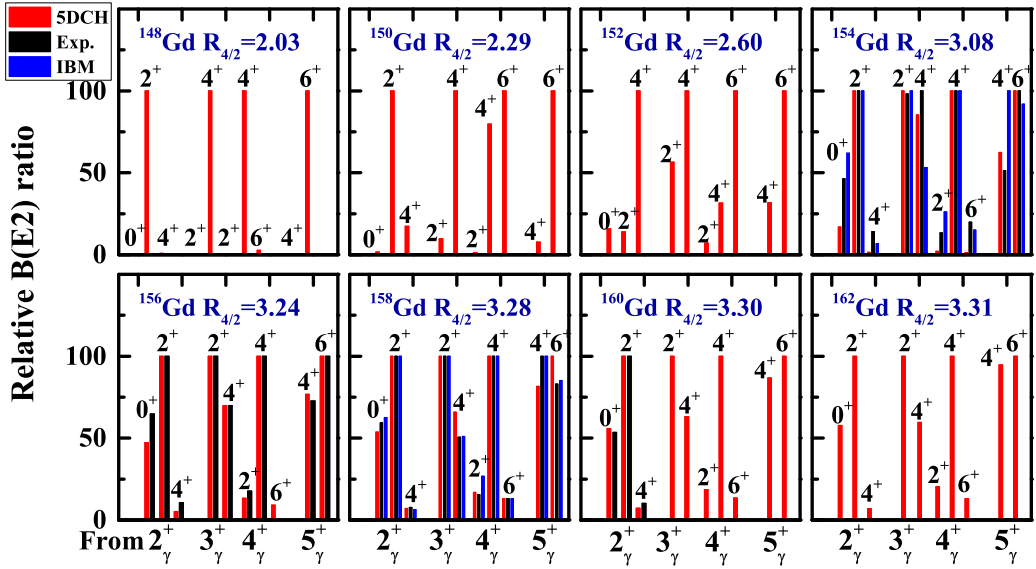}
  \caption{(Color online) Comparison of 5DCH predictions with the data on the relative $\gamma$
  band to ground state band $E2$ transitions in even-even $^{148-162}$Gd isotopes. The red (black)
  bar are the 5DCH predictions (data). The available IBM results \cite{casten2014evidence}
  (blue bar) are also shown for comparison. The largest transition is taken as reference and
  normalized to 100 for each initial state. The $R_{4/2}$ values obtained by 5DCH are also
  given in the figure.}
\label{relative-be2}
\end{figure*}
%-------------------------------------------------------------------------------

The intraband $B(E2)$ transition strength in the g.s., $\gamma$ and $\beta$ bands are also shown
in comparison with the available data in Fig.~\ref{be2}. For $^{148}$Gd, the 5DCH calculation
overestimates the data significantly. This may be because that the excitation mode in $^{148}$Gd does not
belong to the collective excitation, as indicated in Refs.~\cite{podolyak2000multiple,podolyak2003lifetime},
the yrast $0^+$, $2^+$, $4^+$, and $6^+$ states are formed by the coupling of the two $1f_{7/2}$
neutron. The available experimental data in $^{152-162}$Gd are well reproduced by the 5DCH calculations.
It is also found that in $^{154-162}$Gd, the $B(E2)$ transitions of the g.s. and $\beta$ bands are
similar, and the $B(E2)$ values in the $\gamma$ bands are smaller than those of the g.s. and
$\beta$ bands. In addition, the increasing behaviors for the $B(E2)$ of each band in $^{156-162}$Gd
are very similar, indicating similar structures in these four nuclei.

The relative $B(E2)$ ratios from IBM correspond to parameter-free SU(3) PDS calculations and therefore they
are ideal quantities to decide about the validity or not of the concept of SU(3) PDS for a given nucleus or model.
Fig.~\ref{relative-be2} presents the comparison of the relative $B(E2)$ ratios calculated by
5DCH with the available IBM results~\cite{casten2014evidence} and experimental
data~\cite{http://www.nndc.bnl.gov/,casten2014evidence} for Gd isotopes. The theoretical $R_{4/2}$
ratios ($R_{4/2}=E(4_1^+)/E(2^+_1)$) are also shown in the figure. It is found that the available
experimental data display the general characteristic that the spin-decreasing transitions are
larger than the spin-increasing transitions for $^{156-160}$Gd, except for the transition from
$5^+_\gamma$ in $^{156}$Gd, which has an opposite behavior. The spin-conservation transitions
are the largest for transitions from even-spin states in the $\gamma$ bands. The data of
$^{156-160}$Gd can be well reproduced by the 5DCH theoretical calculations. For transitions
from $3^+_\gamma$, $4^+_\gamma$, and $5^+_\gamma$ states in the transitional nucleus $^{154}$Gd,
the spin-increasing transitions are larger than the spin-decreasing ones, and the transition
from $2^+_\gamma$ to $0^+_1$ is larger than that to $4^+_1$. The spin-conservation transitions
are also the largest for the transitions from even spin states in the $\gamma$ band. The
data of $^{154}$Gd can be also reasonably reproduced by the 5DCH calculations.

For the results obtained from 5DCH, it is clearly seen that for $^{148}$Gd, there only exist
spin-conservation transitions from the even-spin states and spin-increasing transitions from
odd-spin states in the $\gamma$ band. The transitions to other states in the g.s. band are so
weak that they are not visible. This can be explained by the vibrator selection rules as emphasized
in Ref.~\cite{couture2015extended}, that the spin-decreasing transitions from non-yrast to yrast
states are forbidden due to the phonon number which can be only changed by 1. Similar phenomena can
be found in $^{150}$Gd. Only the transitions from the $4_\gamma^+$ state in $^{150}$Gd are different
from those in $^{148}$Gd. For transitions from the $4_\gamma^+$ state in $^{150}$Gd, the transition
to $6^+_1$ is the largest, and the spin-conservation transition is a bit small. For $^{152}$Gd,
the relative relations for transitions from $3^+_\gamma$, $4^+_\gamma$, and $5^+_\gamma$
are similar to those in $^{150}$Gd, and the spin-increasing transitions are the largest, while
the transitions from $2_\gamma^+$ in $^{152}$Gd are different from those in $^{150}$Gd. The
transition from $2^+_\gamma$ to $4^+_1$ is much larger than the transition to $2_1^+$, and
the spin-conservation $2^+_\gamma$ to $2^+_1$ transition is the smallest. This behavior can
be understood by the significant mixing of $K$ quantum numbers between $K=0$ and $K=2$ of the
$\gamma$ bandhead for $^{152}$Gd. The corresponding $K$ distributions of the $\gamma$ band
head are 52\%, and 48\% for $K=0$ and $K=2$, respectively. For $^{154}$Gd, the 5DCH spin-decreasing
transitions are larger than the spin-increasing ones, except for the transitions from the $5_\gamma^+$
state, which have an opposite trend. The spin-conservation transitions are the largest for
the even-spin states in the $\gamma$ band. The relative $B(E2)$ transition ratios for $^{156-162}$Gd
are similar to those of $^{154}$Gd, with only small differences in the transition strength.

As shown in Fig.~\ref{relative-be2}, the available IBM results show good agreements with the
data. For $^{158}$Gd, the differences between IBM and 5DCH are smaller than those in $^{154}$Gd.
As described in Ref.~\cite{casten2014evidence}, PDS systematically underestimates the
spin-increasing transitions and overestimates the spin-decreasing transitions in the rare-earth
region. However, the discrepancies between the data and 5DCH results are not systematical. This
could be explained by the fact that the 5DCH calculations have taken the configuration mixing
into account self-consistently. Furthermore, for $^{154-156}$Gd, the calculated transitions
from the $5_\gamma^+$ states by 5DCH are opposite to the PDS descriptions, which predict the
$5_\gamma^+$ to $4^+_1$ transition larger than the $5_\gamma^+$ to $6^+_1$ transition. However,
the transitions from $5_\gamma^+$ states to $6_1^+$ are larger than those to $4^+_1$ states calculated
by 5DCH, which are consistent with the experimental values, e.g., in $^{156}$Gd.

This agreement is not trivial, by including the configuration mixing effect self-consistently,
the parameter-free 5DCH calculations can not only well reproduce the data of relative $B(E2)$
for well-deformed rotors $^{156-162}$Gd, but also reproduce the data for the transitional nucleus
$^{154}$Gd, for which the SU(3) PDS is not expected to work~\cite{casten2014evidence}.
From the 5DCH-CDFT calculations, it seems that $^{156-162}$Gd may fulfil the concept of SU(3) PDS.

%-------------------------------------------------------------------------------
\subsection{Quadrupole deformation}
%-------------------------------------------------------------------------------

Experimentally, with the reduced quadrupole transition probabilities, the deformation parameter
$\beta$ could be extracted~\cite{bohr1975nuclear,ring1980nuclear}. In the framework of 5DCH,
the deformation of a nucleus can be characterized by the average of the invariant $\beta^2$,
$\beta^3\cos3\gamma$ and their combination. Take $\beta^2$ as an example, the average value
of $\beta^2$ in the state $|\alpha I\rangle$ is
%-------------------------------------------------------------------------------
\begin{equation}
    \langle\beta^2\rangle_{I\alpha}=\langle\Psi^I_\alpha|\beta^2|\Psi^I_\alpha\rangle
    =\sum_{K\in\Delta I}\int\beta^2|\psi^I_{\alpha,K}(\beta,\gamma)|^2d\tau_0.
\end{equation}
The expectation values of the quadrupole deformation $\langle\beta\rangle$
and $\langle\gamma\rangle$ together with their fluctuations $\Delta\beta$ and $\Delta\gamma$
for each given state can be obtained \cite{nikvsic2009beyond}
\begin{align}
&\langle\beta\rangle=\sqrt{\langle\beta^2\rangle},\\
&\langle\gamma\rangle=\arccos(\langle\beta^3\cos3\gamma\rangle/\sqrt{\langle\beta^4\rangle\langle\beta^2\rangle})/3,
\end{align}
and
\begin{align}
    \Delta\beta&=\frac{\sqrt{\langle\beta^4\rangle-\langle\beta^2\rangle^2}}{2\langle\beta\rangle},\\
    \Delta\gamma&=\frac{1}{3\sin3\langle\gamma\rangle}\sqrt{\frac{\langle\beta^6\cos^23\gamma\rangle}{\langle\beta^6\rangle}-
    \frac{\langle\beta^3\cos3\gamma\rangle^2}{\langle\beta^4\rangle\langle\beta^2\rangle}}.
\end{align}

%-------------------------------------------------------------------------------
\begin{figure}[h!]
  \centering
  \includegraphics[scale=0.45,angle=0]{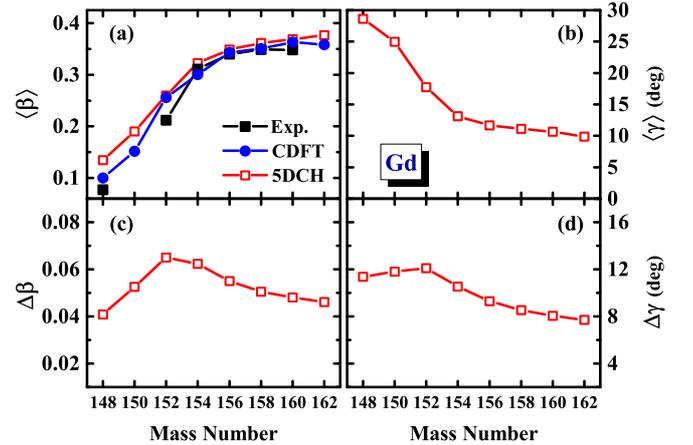}
  \caption{(Color online) The deformation parameters $\langle\beta\rangle$ and $\langle\gamma\rangle$ as
  well as the corresponding fluctuations $\Delta \beta$ and $\Delta \gamma$ of $0_1^+$ state in even-even
  $^{148-162}$Gd isotopes calculated by 5DCH model based on CDFT. The experimental
  data~\cite{http://www.nndc.bnl.gov/} (black cubic) are also
  presented.}
\label{shape-Gd}
\end{figure}
%-------------------------------------------------------------------------------

%-------------------------------------------------------------------------------
\begin{figure}[h!]
  \centering
  \includegraphics[scale=0.45,angle=0]{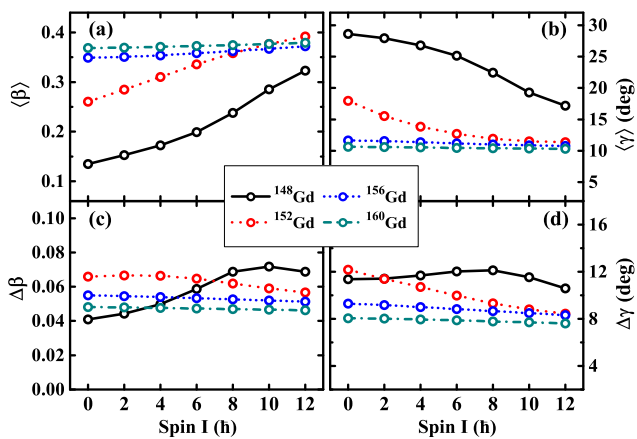}
  \caption{(Color online) The deformation parameters $\langle\beta\rangle$ and $\langle\gamma\rangle$
  as well as the corresponding fluctuations $\Delta \beta$ and $\Delta \gamma$ of ground state bands
  in even-even $^{148, 152, 156, 160}$Gd isotopes as functions of spin calculated by 5DCH-CDFT.}
\label{shape-Gd-spin}
\end{figure}
%-------------------------------------------------------------------------------

%-------------------------------------------------------------------------------
\begin{figure*}[htbp]
 \centering
  \begin{minipage}[t]{0.4\textwidth}
   \includegraphics[scale=0.65]{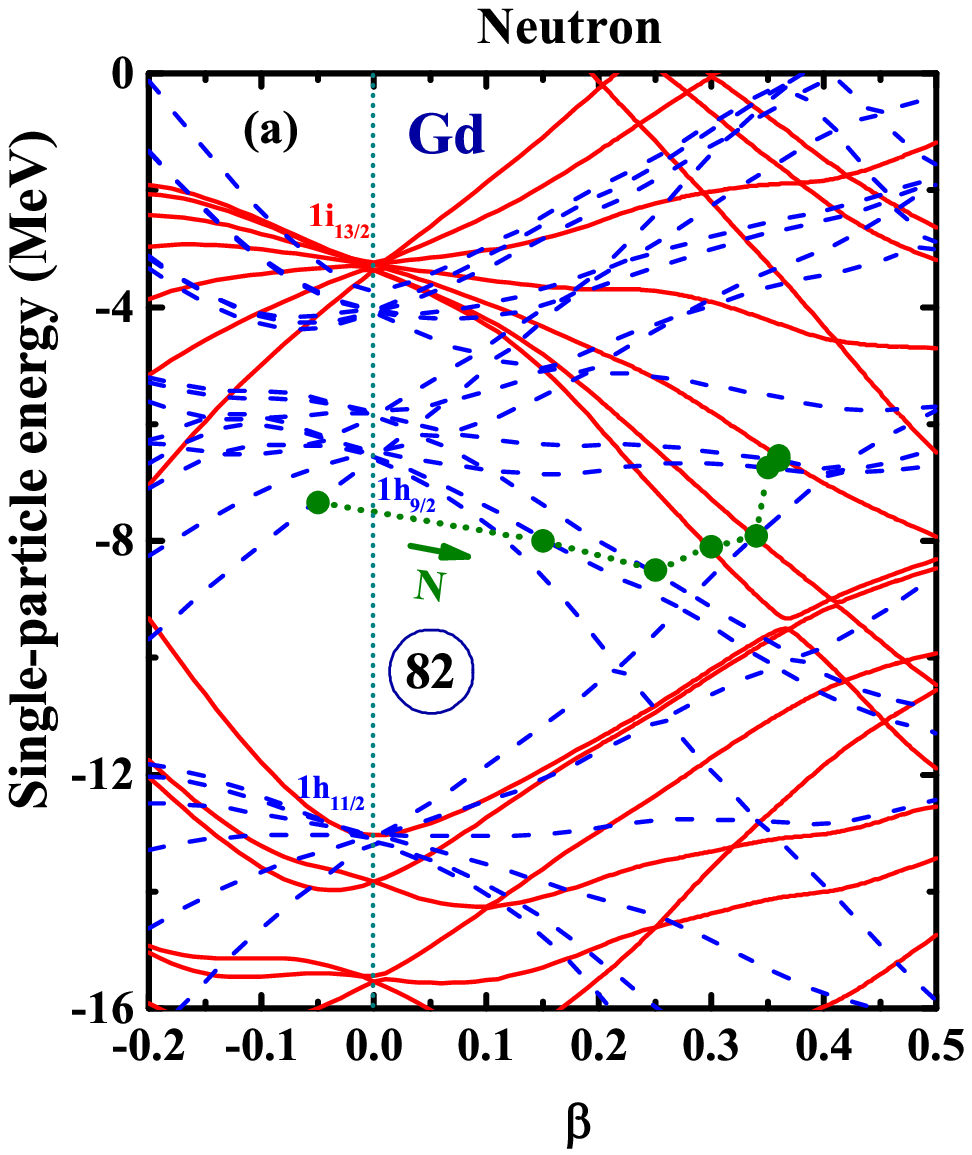}
  \end{minipage}
  \begin{minipage}[t]{0.4\textwidth}
   \includegraphics[scale=0.65]{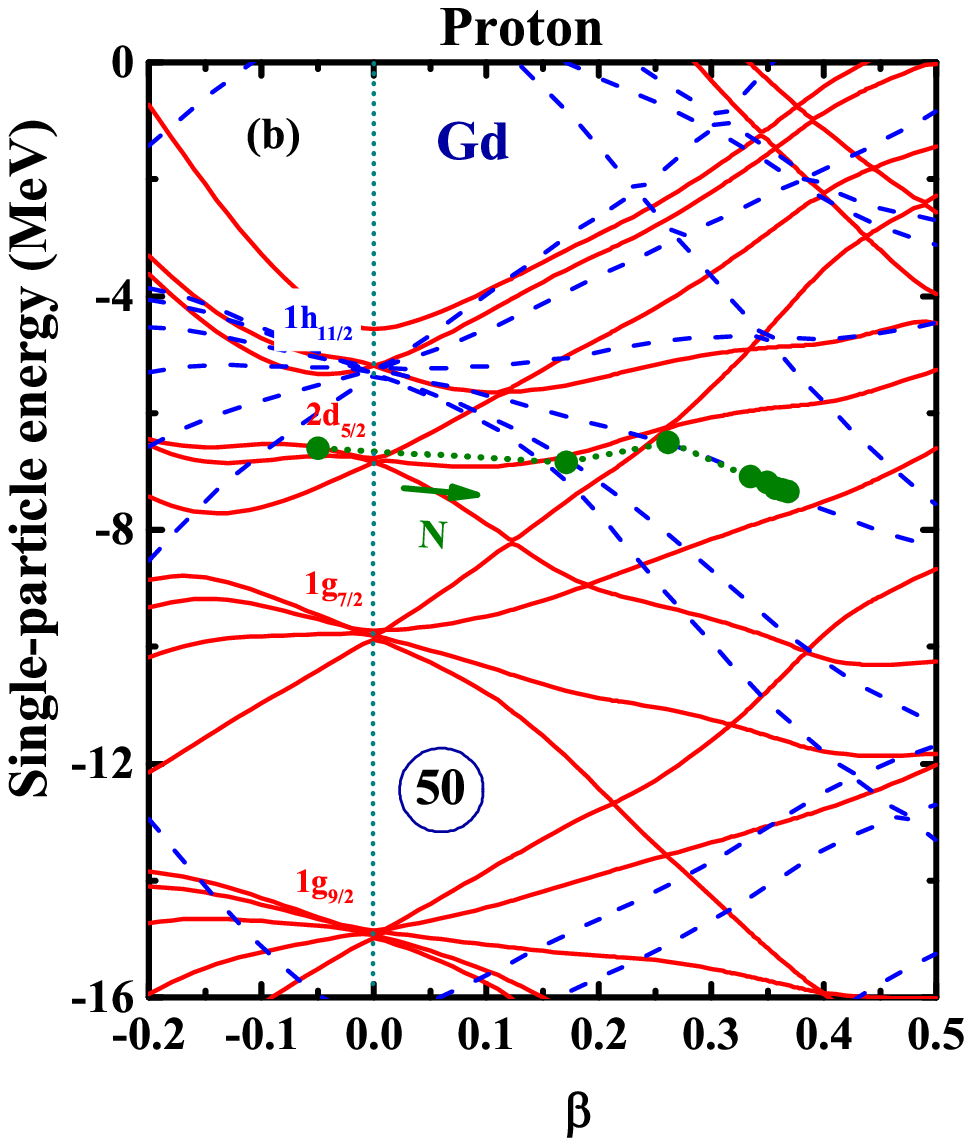}
  \end{minipage}
  \caption{The axially symmetric single neutron (left panel) and proton (right panel) levels of Gd
  as a function of the deformation parameter $\beta$. The red solid (blue dash) curves denote levels
  with positive (negative) parity. The green dots correspond to the positions of the last occupied
  levels at the minimum of the ground states in Gd isotopes.}
\label{single-Gd156}
\end{figure*}
%-------------------------------------------------------------------------------

In Fig.~\ref{shape-Gd}, the obtained results from 5DCH calculations for the ground states of Gd isotopes
are presented. For a comparison, the available experimental data~\cite{http://www.nndc.bnl.gov/}
of $\beta$ deformation are also shown in Fig.~\ref{shape-Gd} (a). It can be clearly seen from
Fig.~\ref{shape-Gd} (a) that with increasing mass number, the $\langle\beta\rangle$ becomes larger
and larger. This is consistent with the results of CDFT as shown in Fig.~\ref{PES-Gd}, where the
$\beta$ values of the minimal of the PES also increase as the mass number increases. The available
experimental data increase with mass number. Clearly, the 5DCH calculations reproduce the data
rather well. For $^{148,152}$Gd, the data are a little overestimated.

The trend of $\langle\gamma\rangle$ is opposite to that of $\langle\beta\rangle$, i.e., decreases
with mass number. The $\langle\gamma\rangle$ starts from nearly $30^\circ$ for $^{148}$Gd to
$10^\circ$ for $^{162}$Gd. It is noted that the $\langle\gamma\rangle$ differences among
$^{154-162}$Gd are quite small, which means that $\langle\gamma\rangle$ has reached a stable
value, and the triaxiality of Gd isotopes becomes stable against the neutron number. Similar
phenomena can also be found in the PES figures. The trends of $\Delta\beta$ and $\Delta\gamma$
for Gd isotopes are similar. The shape fluctuation increases with mass number at first up to
$^{152}$Gd, then begins to decrease. This indicates that $^{152}$Gd is a transitional nucleus
in the even-even $^{148-162}$Gd isotopes.

In Fig.~\ref{shape-Gd-spin}, taking $^{148,152,156,160}$Gd as examples, we plot the
$\langle\beta\rangle$ and $\langle\gamma\rangle$ together with the fluctuations of the quad-rupole
deformations $\Delta\beta$ and $\Delta\gamma$, as functions of spin for the ground state bands.
For $^{148}$Gd, the $\langle\beta\rangle$ value increases with spin, while the $\langle\gamma\rangle$
value decreases with spin rapidly. For its quadrupole fluctuations $\Delta\beta$ and $\Delta\gamma$,
they exhibit firstly increasing and then decreasing trends. The expectation values of
$\langle\beta\rangle$ and $\langle\gamma\rangle$ for $^{152}$Gd are similar to those for $^{148}$Gd,
but with larger a initial value $\langle\beta\rangle=0.26$ and smaller initial value
$\langle\gamma\rangle=18^\circ$. With the increase of spin, $\langle\beta\rangle$ will keep increasing while
$\langle\gamma\rangle$ will finally reach a stable value of $\langle\gamma\rangle\sim12^\circ$.
The strong dependence on the spin of the deformations presented here by $^{148, 152}$Gd is
associated to its PES, shown in Fig.~\ref{PES-Gd}, where the soft behavior can be clearly seen.
For $^{156,160}$Gd, all of the four quantities $\langle\beta\rangle$, $\langle\gamma\rangle$,
$\Delta\beta$, and $\Delta\gamma$ are almost independent of spin. This indicates that the shapes
of $^{156,160}$Gd are so stable that their deformations do not change with spin. Thus, the energy
spectra for $^{156,160}$Gd shown in Figs.~\ref{spectra-gsband},~\ref{spectra-gammaband}, and
~\ref{spectra-betaband} exhibit the rigid rotor behaviors.

\subsection{Single particle levels}

To investigate the shape evolution in Gd isotopes microscopically, the neutron and proton
single-particle levels as functions of the quadrupole deformation $\beta$ in Gd isotopes
obtained from the CDFT are illustrated. As the single particle levels for the Gd isotopes
are similar, here, we only take $^{156}$Gd as an example, as shown in Fig.~\ref{single-Gd156}.
The deformations of the ground states in Gd isotopes are also marked in the last occupied
levels.

For $^{148}$Gd, which possesses a near spherical oblate shape, the last neutron occupies
the $1h_{9/2}$ orbital and the levels around the minimum are very dense in both neutron
and proton single particle levels, which means that the configuration is easy to change.
This corresponds to the soft character around $\beta\sim0.1$ of $^{148}$Gd as shown in
Fig.~\ref{PES-Gd}. For $^{150-152}$Gd, the last neutron still occupies the $1h_{9/2}$
orbital, while the level densities around the minimum point become less dense than $^{148}$Gd.
As a result, the PESs become a bit more rigid. With two more neutrons, the orbital occupied
by the last neutron in $^{154}$Gd changes to the $1i_{13/2}$. As an intruder orbital, the
interaction between the $1i_{13/2}$ and the adjacent negative parity $1h_{9/2}$ levels
is very weak, therefore, the ground state configuration of $^{154}$Gd becomes relatively
stable. With the increase of neutron numbers, more neutrons occupy the $1i_{13/2}$. Thus,
the ground state configurations of $^{156-162}$Gd are difficult to change. Besides, a clear
energy gap could be seen around the last occupied levels in the $^{156-162}$Gd. As a result,
the PESs of $^{156-162}$Gd show well-deformed characters. The occupied $1i_{13/2}$ levels
in the ground states of Gd isotopes make the deformations so stable that even do not change
with the increase of angular momentum, as shown in Fig.~\ref{shape-Gd-spin}.

As for the proton single particle levels, the last occupied level is $2d_{5/2}$ in $^{148}$Gd,
and $1h_{11/2}$ for all the other nuclei. Since the proton numbers for the Gd isotopes are
identical, the difference of the occupation situations is caused by different neutron numbers.
Therefore, the occupations of neutron $1i_{13/2}$ levels, which lead to the well-deformed
prolate shapes of $^{156-162}$Gd and result in the rigid rotor behavior for the energy spectra,
are essential for these isotopes.

%-------------------------------------------------------------------------------
\section{Summary}\label{sec3}
%-------------------------------------------------------------------------------

In conclusion, the low-lying states for the even-even $^{148-162}$Gd isotopes have been
investigated in the framework of 5DCH-CDFT. A clear shape evolution from weakly deformed
$^{148,150}$Gd to $\gamma$-soft $^{152,154}$Gd to well-deformed prolate $^{156-162}$Gd has
been presented. The shapes of $^{156-162}$Gd are all well-deformed prolate with the minima
located at $\beta\sim0.35$, and these deformations are almost independent on the angular
momentum. The available experimental data are reproduced by the calculations, including the
energy spectra, $\gamma$ band staggering and intraband $B(E2)$ transition probabilities
for ground-state $\gamma$ and $\beta$ bands. From the energy spectra, it is found that
the SU(3) symmetry is conserved approximately in the ground state and $\gamma$ bands, while
broken in the $\beta$ bands. Furthermore, by including the configuration mixing self-consistently,
the 5DCH calculations can not only describe the relative $B(E2)$ ratios for the PDS candidates
$^{158}$Gd well, but also reproduce the relative $B(E2)$ ratios for transitional nucleus
$^{154}$Gd. It is found that the occupations of neutron $1i_{13/2}$ orbitals, which lead to
the stable PESs of $^{156-162}$Gd, are essential for these isotopes.

Precious supervision from Prof. Jie Meng and fruitful discussion with Prof. Peter Ring
and Dr. Pengwei Zhao are highly acknowledged. This work was partly supported by the Chinese
Major State 973 Program No.2013CB834400, the National Natural Science Foundation of China
(Grants No. 11375015, 11461141002, 11175002, 11335002), the Research Fund for the Doctoral
Program of Higher Education (Grant No. 20110001110087), Deutsche Forschungsgemeinschaft (DFG) 
and National Natural Science Foundation of China (NSFC) through funds provided to the 
Sino-German CRC 110 "Symmetries and the Emergence of Structure in QCD'' and the China 
Postdoctoral Science Foundation (Grant No. 2015M580007 and No. 2016T90007).

\end{document}